\documentclass[preprint,onecolumn,amsmath,amssymb]{revtex4}
\usepackage{CJK}
\usepackage{graphicx}% Include figure files
\usepackage{dcolumn}% Align table columns on decimal point
\usepackage{bm}% bold math
\usepackage{times}
\usepackage{amsmath}
\usepackage{graphicx}
\usepackage{subfig}
\usepackage{caption}
\usepackage{epstopdf}
\usepackage{color}
\usepackage[abs]{overpic}
\usepackage{natbib}
\begin{document}
\title{ Mean-squared displacement and variance for confined Brownian motion }
%\title{Diffusion behaviors dependent on initial conditions in the confined geometry }
\author{Yi Liao$^{1,}$\footnote{liaoyitianyi@gmail.com}%,%Zi-Kai Fu$^{1}$
and Xiao-Bo Gong$^{2}$}
\affiliation{$^1$ Department of Physics and Information Engineering, College of Mechanical and Electronic Engineering, Fujian Agriculture and Forestry University, Fuzhou, 350002, China\\
$^2$ School of Opto-Electronic Engineering, Zaozhuang University, Zaozhuang, 277160, Shandong, China}
\begin{abstract}
  For one-dimension Brownian motion in the confined system with the size $L$, the mean-squared displacement(MSD) defined by $\left \langle (x-x_0)^2 \right\rangle$ should be proportional to $t^{\alpha(t)}$. The power $\alpha(t)$ should range from $1$ to $0$ over time, and the MSD turns from $2Dt$ to $c L^2$, here the coefficient $c$ independent of $t$, $D$ being the diffusion coefficient. The paper aims to quantitatively solve the MSD in the intermediate confinement regime. The key to this problem is how to deal with the propagator and the normalization factor of the Fokker-Planck equation(FPE) with the Dirichlet Boundaries. Applying the Euler-Maclaurin approximation(EMA) and integration by parts for the small $t$, we obtain the MSD being $2Dt(1-\frac{2\sqrt{\xi} }{3\pi\sqrt{\pi}})$, with $t_{ch}=\frac{L^2}{4\pi^2D},\xi\equiv \frac{t}{t_{ch}}$, and the power $\alpha(t)$  being $\frac{1-0.18\sqrt{\xi}}{1-0.12\sqrt{\xi}}$. Further, we analysis the MSD and the power for the $d$-dimension system with $\gamma$-dimension confinement. In the case of $\gamma< d$, when $t$ is small or large enough, the diffusion is normal($MSD\propto t$). However, there exists the sub-diffusive behavior in the intermediate time. The universal description is consistent with the recent experiments and simulations in the micro-nano systems. Finally, we calculate the position variance(PV) meaning $\left\langle (x-\left\langle   x \right\rangle)^2 \right\rangle$. In the finite system, the variance is not necessarily the same as MSD. The initial conditions are essential to characterize the diffusion behavior described by the FPE, especially in the finite system. Under the initial condition referring to the different probability density function(PDF) being $p_{0}(x)$, MSD and PV should exhibit different dependencies on time, which reflect corresponding diffusion behaviors.As examples, the paper discusses the representative initial PDFs reading $p_{0}(x)=\delta(x-x_0)$, with the midpoint $x_0=\frac{L}{2}$ and the endpoint $x_0=\epsilon$(or $0^+$). In the case of midpoint, the MSD(equal to PV) reads $2Dt(1-\frac{5\pi^3 Dt}{L^2})$ for the small $t$, which reflects a kind of sub-diffusion, with $D$ being the diffusion coefficient. In the case of endpoint, the MSD(equal to PV) reads $\frac{4}{\pi}(2Dt)[1+\frac{2\sqrt{\pi Dt}}{L}]$ for the small $t$, which reflects a kind of super-diffusion.
\end{abstract}
\maketitle

%\newpage
\section*{I.Introduction}
The study of diffusion phenomenon originated from people's exploration of Brownian motion, and its theoretical basis is mainly statistical physics and molecular dynamics\cite{Kardar, Mazur}. At the beginning of the $19$-th century, the British botanist R. Brown found that the suspended small particles such as pollen in the water kept moving in an irregular curve, which was called Brownian motion\cite{Bian, Plyukhin}. Decades later, physicists such as J. Delsaulx, A. Einstein, and P. Langevin et. al. provided a good quantitative explanation for this phenomenon: the mean square displacement (MSD) of small particles is proportional to the observation duration ($MSD\propto t^\alpha,\alpha=1$). Its comprehensive mathematical description corresponds to the probability theory of random walking. Further research has shown that this proportional relationship to the power of time is only applicable to normal diffusion situations. There are also some anomalous diffusion phenomena in nature, such as sub diffusion and super-diffusion. $\alpha=0$ corresponds to strict localization, and $\alpha=2$ corresponds to ballistic transport, which corresponds to the power relationship of uniform motion. The transition between localization and normal diffusion is called sub-diffusion, while the transition between normal diffusion and ballistic transport is called super-diffusion. The extended diffusion model can explain many phenomena in physics, chemistry, biology, virus transmission, and even economic activities\cite{Dzugutov,de Grooth,Plyukhin2006,Liao2021}.

For the diffusion, researchers mainly consider the transport properties of their internal properties, with little exploration of the influence of boundary conditions on them, generally limited to free infinite space or periodic boundary conditions. However, the confinement effect require more critical and cautious treatment. For example, in the Brownian motion in a cup, as time increases, the square root of the mean square displacement of particles cannot exceed the physical scale $\rho$ of the cup. After a sufficient period of time, the mean square displacement of particles is only related to and the dependence on time gradually disappears($MSD\propto t^{\alpha(t,\rho)},\alpha(t,\rho):1\rightarrow 0 $). This phenomenon naturally goes against the rule that the MSD is proportional to the observation duration. The quantitative description of this intuitive feeling also has academic appeal and considerable scientific significance. With the refinement and deepening of research, this confinement effect can essentially be attributed to the influence of scale effects, and its importance will be highlighted in low dimensional and microscale situations \cite{Faucheux,Alonso,Krager,Liao2015,Ernst}.

 The confinement has been shown to the sub-diffusive dynamics of particles and macromolecules in  micro-nano system, special the biological system\cite{Hitimana,Aporvari}.
 Several studies have reported the sub-diffusion behaviors in confining system, such as the slits, spheres, channels, and other geometries\cite{Broersma,Pawar,Bevan,Lin,Kazoe}.The kind of slowdown was more pronounced as the degree of confinement increased. However, the previous papers have rarely explored the confinement effect purely from the perspective of boundary conditions, but have focused more on the size effect through  comparing the Brownian particle scale with the confinement scale and exploring the effective diffusion coefficient. They avoid the tedious task of normalizing the conservation of probability in finite space. The paper attempts to study the confinement effect from the viewpoint of normalization factor. The normalization factor is equivalent to the partition function in statistical physics, and many confinement effects can be attributed to this. For example, the crucial mean-first-passage time(MFPT) in heat conduction problems can be considered as the integration with time variables of the time-dependent normalization factors.
\section*{II. Propagator and normalization factor in finite system}
 The propagator satisfies the Fokker-Planck Equation in the confined system\cite{Gitterman}.
\begin{equation}
\frac{\partial}{\partial t} Q(x,t|x_0,0)=(-F\frac{\partial}{\partial x}+D \frac{\partial^2}{\partial x^2})Q(x,t|x_0,0) .
\end{equation}
Dirichlet boundaries mean
\begin{equation}
Q(0,t|x_0,0)=Q(L,t|x_0,0)=0.
\end{equation}
The corresponding propagator reads
\begin{eqnarray}
Q(x,t|x_0,0)&=&\frac{2}{L}\exp[\frac{2F(x-x_0)-F^2 t}{4D}]
\\ \nonumber &\times&\underset{n=1}{\overset{+\infty}{\sum}}\exp[-\frac{n^2\pi^2D t}{L^2}]\sin(\frac{n\pi x_0 }{L})\sin(\frac{n\pi x }{L}).
\end{eqnarray}
If the external force $F=0$, the propagator reads
\begin{eqnarray}
Q(x,t|x_0,0)&=&\frac{2}{L}\underset{n=1}{\overset{+\infty}{\sum}}\exp[-\frac{n^2\pi^2D t}{L^2}]\sin(\frac{n\pi x_0 }{L})\sin(\frac{n\pi x }{L}).
\end{eqnarray}
To analysis the diffusion behavior,we need to know the probability density function $P(x,t)$. To keep probability conserved, we have to obtain the normalization factor in different initial condition. In the paper, we discuss three initial condition.
\\Initial condition \textcircled{1} means
\begin{equation}
Q(x,0|x_0,0)=\delta(x-x_0),p_0(x_0)=\frac{1}{L}.
\end{equation}
Here, the $p_0(x_0)=\frac{1}{L}$  denotes the uniform distribution for the initial point.
The PDF reads
\begin{equation}
P(x,t)=\frac{1}{L \bar{Z}(L,t)} \int_0^L Q(x,t|x_0,0)dx_0, \int Q(x,t|x_0,0)dx_0 dx=L\bar{Z}(L,t).
\end{equation}
The normalization factor $\bar{Z}(L,t)$ reads
\begin{equation}
\bar{Z}(L,t)=\frac{8}{\pi^2}\underset{m=0}{\overset{+\infty}{\sum}}\exp[-A(2m+1)^2]\frac{1}{(2m+1)^2}, A  \equiv \frac{\pi^2Dt }{L^2}.
\end{equation}
The mean first passage time\cite{Gitterman,Li2003}reads
\begin{eqnarray}
T &=&\int_0^{+\infty} \int_0^L \int_0^L   Q(x,t|x_0,0) dxdx_0dt=\int_0^{+\infty} \bar{Z}(L,t) dt
\\ \nonumber &=& \frac{8L^2}{\pi^4 D}\underset{m=0}{\overset{+\infty}{\sum}}\frac{1}{(2m+1)^4}=\frac{L^2}{12D}=\int_0^L [\frac{x_0(L-x_0)}{2D}]p_0(x_0)dx_0.
\end{eqnarray}
\\Initial condition \textcircled{2} reads
\begin{equation}
Q(x,0|x_0,0)=p_0(x)=\delta(x-\frac{L}{2}).
\end{equation}
The normalization factor $Z(L,t)$ reads
\begin{eqnarray}
 Z(L,t)&=&\frac{2}{L}\int_0^{L}\underset{m=0}{\overset{+\infty}{\sum}}\{\exp[-\frac{(2m+1)^2\pi^2D t}{L^2}]
\\ \nonumber &\times &[(-1)^m]\sin[\frac{(2m+1)\pi x}{L}]\}dx.
\\ \nonumber &=&\frac{4}{\pi}\underset{m=0}{\overset{+\infty}{\sum}}[\frac{(-1)^m}{2m+1}]\exp[-\frac{(2m+1)^2\pi^2D t}{L^2}].
\end{eqnarray}
The PDF reads
\begin{eqnarray}
P(x,t)&=&\frac{2}{ L Z(L,t)}\underset{m=0}{\overset{+\infty}{\sum}}\{\exp[-\frac{(2m+1)^2\pi^2D t}{L^2}]
\\ \nonumber &\times &[(-1)^m]\sin[\frac{(2m+1)\pi x}{L}]\}.
\end{eqnarray}
\\Initial condition \textcircled{3} reads
\begin{equation}
Q(x,0|x_0,0)=p_0(x)=\delta(x-\epsilon).
\end{equation}
The normalization factor $Z(L,t)$ reads
\begin{eqnarray}
Z(L,t)&=&\frac{2\pi \epsilon}{ L^2 }\int_0^{L}\underset{n=0}{\overset{+\infty}{\sum}}\{\exp[-\frac{n^2\pi^2D t}{L^2}]n\sin[\frac{n\pi x}{L}]\}dx
\\ \nonumber &=&\frac{4 \epsilon}{  L }\underset{m=0}{\overset{+\infty}{\sum}}\exp[-\frac{(2m+1)^2\pi^2D t}{L^2}]\equiv \frac{4 \epsilon}{  L } \hat{Z}(L,t).
\end{eqnarray}
The PDF reads
\begin{eqnarray}
P(x,t)=\frac{\pi}{ 2L \hat{Z}(L,t)}\underset{n=0}{\overset{+\infty}{\sum}}\{\exp[-\frac{n^2\pi^2D t}{L^2}]n\sin[\frac{n\pi x}{L}].\}
\end{eqnarray}
To deal with all kinds of sums of series, we introduce the Euler-Maclaurin approximation(EMA) which means
\begin{eqnarray}
\sum_{m=0}^{\infty}M(m)&\approx& \int_{0}^{+\infty}M(x)dx +\frac{M(0)+M(+\infty)}{2}\nonumber\\
&+& \sum_{k=1}^{\infty}\frac{B_{2k}}{(2k)!}[\frac{d^{(2k-1)}M}{dx^{(2k-1)}}(+\infty)-\frac{d^{(2k-1)}M}{dx^{(2k-1)}}(0)].
\label{eq:EMA}
\end{eqnarray}
\begin{figure}
 \centering
   \includegraphics[width=0.8 \textwidth]{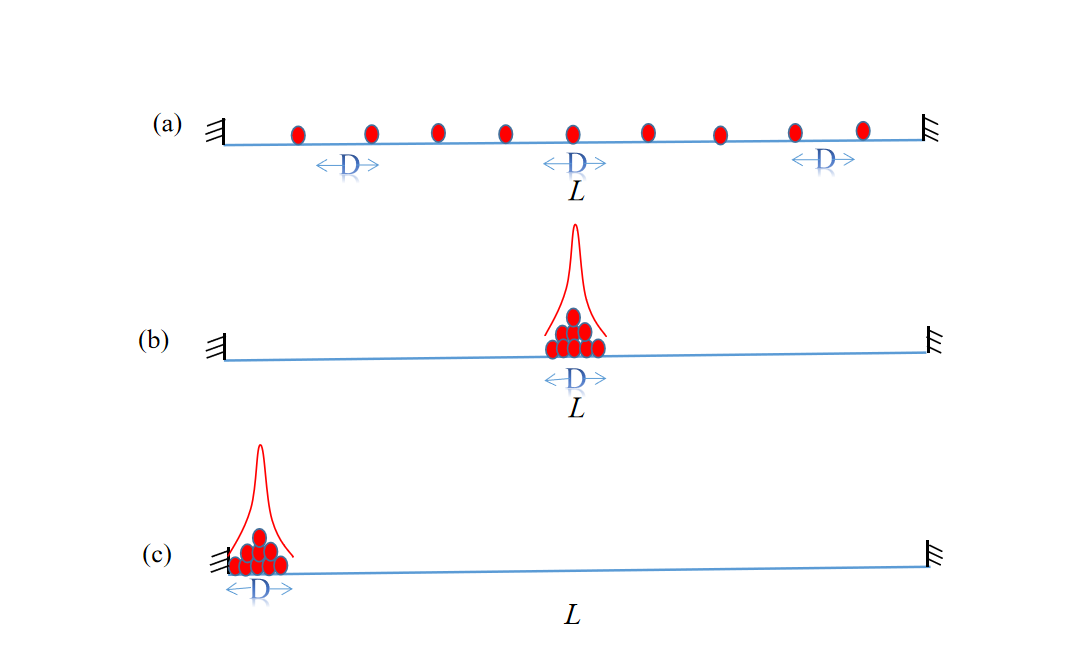}
   \caption{Diagrammatic sketch for the initial condition:(a),(b) and (c) corresponding \textcircled{1},\textcircled{2} and\textcircled{3}. The symbol "D" denotes diffusion behavior and the symbol "$L$" denotes the confined size.}
   \label{fig:1}
\end{figure}
\section*{III. MSD and PV for confined Brownian motion}
\label{sect��MSD}
\begin{figure}
 \centering
   \includegraphics[width=0.8 \textwidth]{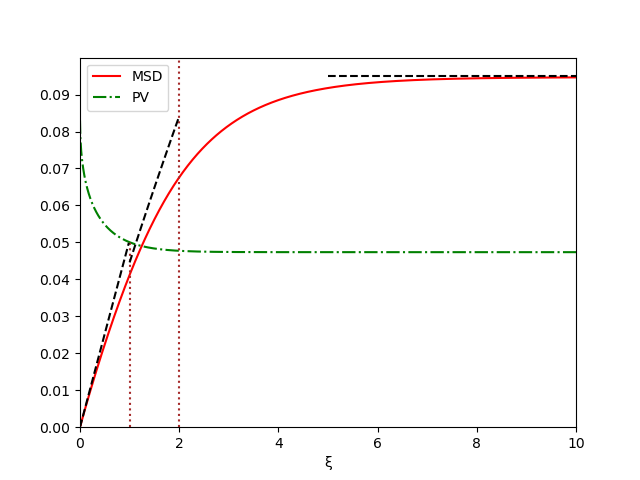}
   \caption{Numerical result of MSD and PV (in the unit of $L^2$) dependent of the reduced time $\xi$($\xi=t/t_{ch},t_{ch}=\frac{L^2}{4\pi^2D}$) for initial condition (a) through summing of $100$ terms. The black dashed line denotes the corresponding analytic result shown in Table.\ref{Tab:1}, with the asymptotic value $0.095$.}
   \label{fig:2}
\end{figure}

The mean square displacement(MSD)for the  one-dimension system in initial condition \textcircled{1}  is defined by
\begin{equation}
\left\langle   (x-x_0)^2 \right\rangle=\frac{1}{ L \bar{Z}(L,t)}\int (x-x_0)^2Q(x,t|x_0,0)dx_0dx.
\end{equation}
The probability density function(PDF) reads
\begin{eqnarray}
P(x,t)&=&\frac{4}{\pi L \bar{Z}(L,t)}\underset{m=0}{\overset{+\infty}{\sum}}\{\exp[-\frac{(2m+1)^2\pi^2D t}{L^2}]
\\ \nonumber &\times &[\frac{1}{(2m+1)}]\sin[\frac{(2m+1)\pi x}{L}]\}.
\end{eqnarray}
We re-write the normalization factor reads\cite{Liao2015}
\begin{equation}
\bar{Z}(L,t)=\frac{8}{\pi^2}\underset{m=0}{\overset{+\infty}{\sum}}\exp[-A  (2m+1)^2]\frac{1}{(2m+1)^2}\equiv \frac{8}{\pi^2} I(t).
\end{equation}
we introduce the reduced size meaning $\tilde{L}^{-1}\equiv \sqrt{A}=\frac{\pi \sqrt{D t} }{L}.$ If $\tilde{L}^{-1}$ is small, Using the Eq.(\ref{eq:EMA}), we obtain
\begin{equation}
 I(t)=\frac{\pi^2}{8}[1-\frac{4}{\pi^{3/2}}\tilde{L}^{-1}+O(\tilde{L}^{-4})]\approx\frac{\pi^2}{8}[ 1-\frac{4}{\sqrt{\pi}}\frac{1}{L}\sqrt{D t}]+O(t^2).
 \label{eq:I}
\end{equation}
The normalization factor reads
\begin{equation}
\bar{Z}(L,t)=1-\frac{4}{\pi^{3/2}}\tilde{L}^{-1}+O(\tilde{L}^{-4})\approx 1-\frac{4}{\sqrt{\pi}}\frac{\sqrt{D t}}{L}.
\end{equation}
Making the Taylor expansion of normalization factors, we have
\begin{equation}
1/\bar{Z}(L,t)=1+4\beta t^{1/2}+16\beta^2t+64\beta^3t^{3/2},\beta\equiv \frac{\sqrt{D }}{\sqrt{\pi}L}.
\end{equation}
The normalization factor is related to the fluctuation-induction force, We have proved that the EMA is effective when $\tilde{L}^{-1}<0.5$ in Ref.\cite{Liao2015}. Ones know $\left\langle x_0^2 \right\rangle=\left\langle   x^2 \right\rangle=\int x^2P(x,t)dx$ in this case. So ones get $\left\langle   (x-x_0)^2 \right\rangle=2[\left\langle   x^2 \right\rangle-\left\langle   x x_0 \right\rangle]$.
Here, the position correlation function reads
\begin{eqnarray}
\left\langle   xx_0 \right\rangle&=&\frac{2L^2}{\pi^2 \bar{Z}(L,t)}\underset{m=0}{\overset{+\infty}{\sum}}\{\exp[-\frac{(m+1)^2\pi^2D t}{L^2}][\frac{1}{(m+1)^2}] \}
\\ \nonumber &\equiv &\frac{2L^2}{\pi^2 \bar{Z}(L,t)} II(t).
\end{eqnarray}
The average of the square of position variable reads
\begin{eqnarray}
\left\langle   x^2 \right\rangle&=&\frac{4L^2}{\pi^4 \bar{Z}(L,t)}\underset{m=0}{\overset{+\infty}{\sum}}\{\exp[-\frac{(2m+1)^2\pi^2D t}{L^2}]
\\ \nonumber &\times &[\frac{1}{(2m+1)^4}][\pi^2(2m+1)^2-4] \}.
\label{eq:x2}
\end{eqnarray}
It can turn into the following formula, which reads
\begin{eqnarray}
\left\langle   x^2 \right\rangle&=&\frac{L^2}{2}-\frac{16L^2}{\pi^4 \bar{Z}(L,t)}\underset{m=0}{\overset{+\infty}{\sum}}\{\exp[-\frac{(2m+1)^2\pi^2D t}{L^2}]\frac{1}{(2m+1)^4}] \}
\\ \nonumber &\equiv &\frac{L^2}{2}-\frac{16L^2}{\pi^2 \bar{Z}(L,t)}III(t).
\end{eqnarray}
We introduce the function $II'(t)$ which reads
\begin{equation}
II'(t)\equiv\frac{6}{\pi^2} II(t)\approx 1-6\beta t^{1/2}+3\pi \beta^2 t.
\end{equation}
Here,the function $III(t)$ satisfies
\begin{eqnarray}
\frac{\partial III(t)}{\partial t}=-\frac{D}{L^2}I(t),III(0)=\frac{\pi^2}{96}.
\end{eqnarray}
We introduce the function $III'(t)$ which reads
\begin{eqnarray}
III'(t)\equiv \frac{96}{\pi^2}III(t)=1-12\pi \beta^2t+32 \pi \beta^3t^{3/2}.
\end{eqnarray}
So, the MSD expressed as a series solution reads
\begin{equation}
\left\langle   (x-x_0)^2 \right\rangle=L^2\{1-\frac{1}{3\bar {Z}(t)}[III'(t)+2II'(t)]\}\equiv L^2 f(t).
\label{eq:MSD}
\end{equation}
Adopting the EMA for the small $t$, we have
\begin{equation}
\left\langle   (x-x_0)^2 \right\rangle=2\pi\beta^2t[1-\frac{4\beta }{3}t^{1/2}]=2Dt(1-\frac{4\sqrt{Dt} }{3\sqrt{\pi}L}).
\label{eq:compare}
\end{equation}
We introduce the characteristic time $t_{ch}=\frac{L^2}{4\pi^2D}$ and the reduced time $\xi\equiv \frac{t}{t_{ch}}. $ The MSD reads
\begin{equation}
\left\langle   (x-x_0)^2 \right\rangle=2Dt(1-\frac{2\sqrt{\xi} }{3\pi\sqrt{\pi}})=\frac{L^2}{2\pi^2}(\xi-0.12\xi^{3/2}).
\end{equation}
And for the large $t$, we can adopt the first-term approximation(FTA) for the series solution in Eq.({\ref{eq:MSD}}). When $A\equiv0.25\xi$ is large (meaning $A>A_0$), the structure factor reads\cite{Liao2015}
\begin{equation}
S(q)\equiv \left\langle   \exp[iq(x-x_0)] \right\rangle=\frac{\pi^4[1+\cos(qL)]}{2(\pi^2-q^2L^2)^2}.
\end{equation}
The MSD reads
\begin{equation}
MSD=-\frac{d^2S(q)}{dq^2}|_{q=0}=\frac{(\pi^2-8)}{2\pi^2}L^2,t\rightarrow +\infty.
\end{equation}
As the shown in Fig.(\ref{fig:2}), $A_0\approx 2.5$, the approximation is reasonable.

We define the power $\alpha(t)$ by\cite{Ernst}
\begin{equation}
\underset{\Delta t\rightarrow 0}\lim \frac{MSD(t+\Delta t)}{2D_{eff}(t+\Delta t)^{\alpha(t)}}=1.
\end{equation}
We have
\begin{equation}
\alpha(t)\equiv \alpha(\xi)=\frac{1-0.18\sqrt{\xi}}{1-0.12\sqrt{\xi}}.
\end{equation}
we know $\alpha(0.5)=0.95,\alpha(1)=0.93,\alpha(2)\approx0.90$. Because $\frac{2}{3\pi \sqrt{\pi}}\approx 0.120$,$\frac{(\pi^2-8)}{2\pi^2}=0.095$, we notice
\begin{equation}
t=2.28t_{ch}, 2Dt(1-\frac{4\sqrt{Dt} }{3\sqrt{\pi}L})=\frac{(\pi^2-8)}{2\pi^2}L^2.
\end{equation}

Above results are Summarized in Table.\ref{Tab:1}.
\begin{table}
\caption{1.The mean-squared displacement$\left\langle (x-x_0)^2 \right\rangle$ and the power $\alpha(t)$ for the one-dimension confined system.$\frac{(\pi^2-8)}{2\pi^2}=0.095$}
\begin{centering}
\begin{tabular*}{12cm}{c|ccccccc}
\hline
\hline
$\xi=t/t_{ch}$ & $A$  & & MSD=$\left\langle   (x-x_0)^2 \right\rangle$ & & $\alpha(t)$ \\
\hline
\hline
$[0,1)$  & $(0,0.25)$ & &$2Dt$& & $1$   \\
\hline
$[1,2)$  & $[0.25,0.5]$&  &$2Dt[1-0.12\sqrt{\xi}]$ & & $\frac{1-0.18\sqrt{\xi}}{1-0.12\sqrt{\xi}}$   \\
\hline
$[2,10)$  & $[0.5,2.5]$&  &$L^2f(t)$ & & $\frac{t}{f(t)}\frac{df}{dt}$   \\
\hline
$[10,+\infty)$  & $(2.5,+\infty)$& & $\frac{(\pi^2-8)}{2\pi^2}L^2=3.74 Dt_{ch}$ & & $0$ \\
\hline
\hline
\end{tabular*}
\end{centering}
\label{Tab:1}
\end{table}
Further, we analysis the MSD and the power for the $d$-dimension system with $\gamma$-dimension confinement. As shown in Table.\ref{Tab:d}, in the case of $\gamma<d$, when $t$ is small or large enough, the diffusion is normal($MSD\propto t$). The fator $\eta_1\approx 1$, $\eta_2\approx 1$ is dependent of the numerical result. The function $g(t)$ is a series summation similar to $f(t)$.
\begin{table}
\caption{2. The mean-squared displacement$\left\langle   (\vec{r}-\vec{r}_0)^2 \right\rangle$ and the power $\alpha(t)$ for the $d$-dimension system with $\gamma$-dimension confinement, $C_{d}^{\gamma}=\frac{2\gamma}{3\pi\sqrt{\pi}d}\approx \frac{0.12\gamma}{d}.$ }
\begin{centering}
\begin{tabular*}{12cm}{c|ccccccc}
\hline
\hline
$\xi=t/t_{ch}$  & & MSD=$\left\langle(\vec{r}-\vec{r}_0)^2 \right\rangle$ & & $\alpha(t)$ \\
\hline
\hline
$[0,1)$  & &$2^dDt$& & $1$   \\
\hline
$[1,\frac{2\eta_1 d^2}{\gamma^2})$  &  &$2^dDt[1-C_{d}^{\gamma}\sqrt{\xi}]$ & & $\frac{1-1.5C_{d}^{\gamma}\sqrt{\xi}}{1-C_{d}^{\gamma}\sqrt{\xi}}<1$   \\
\hline
$[\frac{2\eta_1 d^2}{\gamma^2},\frac{10\eta_2 d^2}{\gamma^2})$  &  &$2^dDt[1-g(t)]$ & & $1-\frac{t}{1-g(t)}\frac{dg}{dt}<1$   \\
\hline
$[\frac{10\eta_2 d^2}{\gamma^2},\infty)$  & & $\frac{\gamma(\pi^2-8)}{2\pi^2}L^2+2^{d-\gamma} Dt$ & & $1(\gamma< d),0(d=\gamma)$ \\
\hline
\hline
\end{tabular*}
\end{centering}
\label{Tab:d}
\end{table}

Using the Eq.(\ref{eq:x2})and considering $\left\langle   x \right\rangle=\frac{L}{2}$, the position variance(PV) reads
\begin{eqnarray}
 PV&\equiv&\left\langle   x^2 \right\rangle-\left\langle   x \right\rangle^2 =\frac{L^2}{4}-\frac{L^2}{6 \bar{Z}(L,t)}III'(t)
\\ \nonumber &\approx& \frac{L^2}{4}-\frac{L^2}{6(1-4\beta t^{1/2})}+2Dt(\frac{1-\frac{8}{3}\beta t^{1/2}}{1-4\beta t^{1/2}}).
\\ \nonumber &\approx& \frac{2L^2}{3}[(3\pi-4)\beta^2t-\beta t^{1/2}]+ \frac{L^2}{12}.
\label{eq:PV}
\end{eqnarray}
\section*{IV. Diffusion behavior dependent on initial condition in the confined geometry }
The initial conditions are essential to characterize the diffusion behavior described by the Fokker-Planck equation, especially in the finite system. To clarify this dependency, we consider the one-dimension FPE with the Dirichlet Boundaries in the confined geometry with the size $L$. Under the initial condition referring to the different probability density function(PDF) being $p_{0}(x)$, the mean-squared displacement defined by $\left\langle (x-x_0)^2 \right\rangle$ and the position variance(PV) meaning $\left\langle (x-\left\langle x \right\rangle)^2 \right\rangle$ should exhibit different dependencies on time, which reflect corresponding diffusion behaviors. The key to this problem is how to deal with the propagator of FPE and the normalization factor. For the small $t$, we also apply the Euler-Maclaurin approximation and integration by parts.
\subsection*{i. Midpoint case}
In midpoint case, the normalization reads
\begin{eqnarray}
 Z(L,t)=\frac{4}{\pi}\underset{m=0}{\overset{+\infty}{\sum}}[\frac{(-1)^m}{2m+1}]\exp[-(2m+1)^2A].
\end{eqnarray}
It also reads
\begin{eqnarray}
 Z(L,t)&=&\frac{4}{\pi}\underset{m=0}{\overset{+\infty}{\sum}}\{(\frac{1}{4m+1})\exp[-(4m+1)^2A ]
\\ \nonumber &- &(\frac{1}{4m+3})\exp[-(4m+3)^2A ]\}.
\\ \nonumber &=&\frac{4}{\pi}\{\int_1^{3}\frac{\exp(-Ax^2)}{x}dx+\frac{1}{2}[\exp(-A )-\exp(-9A)]+\cdots \}
\\ \nonumber &=&\frac{2}{\pi}\{[Ei(-9A)-Ei(-A)+\exp(-A )-\exp(-9A)]+\cdots \}
%\\ \nonumber &+ &\sum_{k=1}^{\infty}\frac{2B_{2k}}{(2k)!}[\exp(-A )-\frac{\exp(-9A )}{3^{2k}}]\}
\\ \nonumber &\approx&1-\frac{40}{\pi} A^2+o(A^2).
\end{eqnarray}
Here, the Airy function reads
\begin{equation}
Ei(\zeta)\equiv \int_{-\infty}^\zeta\frac{\exp(t)}{t}dt=\gamma+ln|\zeta|+ \underset{m=1}{\overset{+\infty}{\sum}}\frac{\zeta^m}{m m!}.
\end{equation}
The average of the square of position variable reads
\begin{eqnarray}
\left\langle    x^2 \right\rangle&=&\frac{2L^2}{ Z(L,t)}\underset{m=0}{\overset{+\infty}{\sum}}\{\exp[-A(2m+1)^2]
\\ \nonumber &\times &[(-1)^m]\frac{(2m+1)^2\pi^2-4 }{(2m+1)^3\pi^3}\}.
\end{eqnarray}
When $A\rightarrow +\infty$, $\left\langle  x^2 \right\rangle=\frac{\pi^2-4}{2\pi^2}L^2$. Considering $\left\langle x \right\rangle=x_0=\frac{L}{2}$, we have
\begin{equation}
MSD(t\rightarrow +\infty)=(\frac{\pi^2-8}{4\pi^2})L^2\approx 0.047 L^2.
\end{equation}

We introduce the auxiliary function $R(A)$, which reads
\begin{equation}
R(A)\equiv \underset{m=0}{\overset{+\infty}{\sum}}\{\frac{[(-1)^m] \exp[-A(2m+1)^2]}{(2m+1)^3}\}
\end{equation}
It satisfies
\begin{equation}
\frac{\partial R(A)}{\partial A}=-\frac{\pi}{4} Z(L,t), R(0)=\frac{\pi^3}{32}.
\end{equation}
Thus,  we get
\begin{eqnarray}
\left\langle   x^2 \right\rangle&=&\frac{L^2}{2}-\frac{8L^2}{\pi^3 Z(L,t)}\underset{m=0}{\overset{+\infty}{\sum}}\{\frac{[(-1)^m] \exp[-A(2m+1)^2]}{(2m+1)^3}\}
\\ \nonumber &\approx& \frac{L^2}{2}-\frac{L^2[1-\frac{8A}{\pi^2}+O(A^3)]}{ 4[1-\frac{40}{\pi}A^2+o(A^2)]}.
\end{eqnarray}
We have the following formula which reads(with small $t$)
\begin{equation}
MSD=PV\approx2Dt(1-\frac{5A}{\pi} )=2Dt(1-\frac{5\pi Dt}{L^2}).
\end{equation}
\subsection*{ii. Endpoint case}
\begin{figure}
 \centering
  \subfloat{\includegraphics[width=0.8 \textwidth]{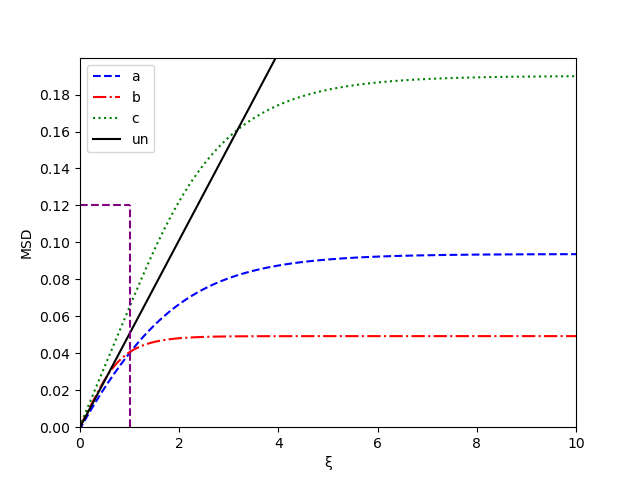}}

  \subfloat{\includegraphics[width=0.8 \textwidth]{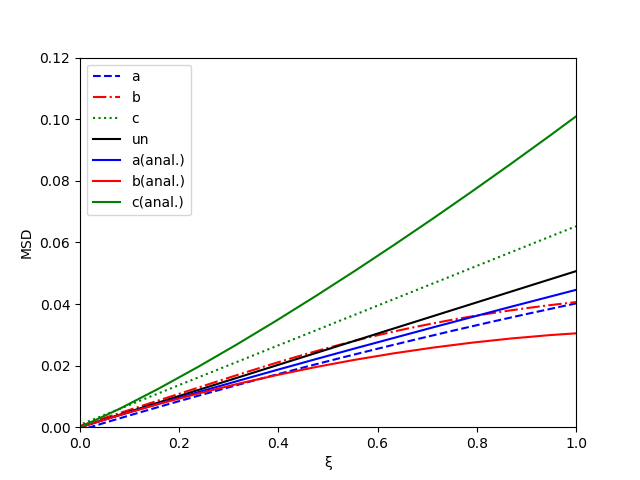}}
   \caption{Top panel: the theoretical result of MSD(in the unit of $L^2$) dependent of $\xi$ for the initial condition (a), (b) and (c) through summing of $200$ terms. The blank solid line labelled by the symbol "un" corresponds the unconfined system where $MSD=2Dt$. The corresponding asymptotic value is $0.095$, $0.047$ and $0.189$. Bottom panel: the comparison between theoretical result and analytical one dependent of small $\xi$. The "anal." means the analytic result. }
 \label{fig:3}
\end{figure}
We have defined the co-error function $erfc(\zeta)$, and for small $\zeta$
\begin{eqnarray}
erfc(\zeta)&\equiv& 1-erf(\zeta) \equiv 1-\frac{2}{\sqrt{\pi}}\int_0^\zeta\exp(-t^2)dt
\\ \nonumber &=& 1-\frac{2}{\sqrt{\pi}}(\zeta-\frac{\zeta^3}{3}+\frac{\zeta^5}{5\cdot 2!}+\cdots).
\end{eqnarray}
Using the EMA, we have
\begin{eqnarray}
\hat{Z}(L,t)&=&{\underset{m=0}{\overset{+\infty}{\sum}}\exp[-(2m+1)^2A]}\approx \frac{\sqrt{\pi}}{4\sqrt{A}} erfc(\sqrt{A}) \sim \frac{\sqrt{\pi}}{4\sqrt{A}}(1-\frac{2}{\sqrt{\pi}}\sqrt{A}).
\label{eq:div}
\end{eqnarray}
Thus,we have
\begin{eqnarray}
MSD=PV&=&\left \langle   x^2 \right\rangle= \frac{L^2 \underset{n=1}{\overset{+\infty}{\sum}}[\frac{(2-n^2\pi^2)(-1)^n-2}{\pi^3n^2}]\exp[-n^2A]}{\underset{m=0}{\overset{+\infty}{\sum}}\exp[-(2m+1)^2A]}
\\ \nonumber &=&\frac{L^2}{\pi \hat{Z}(L,t)} \underset{m=0}{\overset{+\infty}{\sum}}\{\exp[-(2m+1)^2A]-\exp[-(2m+2)^2A]\}
\\ \nonumber &-&\frac{4L^2}{\pi^3 \hat{Z}(L,t)} \underset{m=0}{\overset{+\infty}{\sum}}\frac{\exp[-A(2m+1)^2]}{(2m+1)^2}.
\\ \nonumber &\equiv& \frac{L^2V(t)}{\pi \hat{Z}(L,t)} -\frac{4L^2I(t)}{\pi^3 \hat{Z}(L,t)}.
\end{eqnarray}
When $A\rightarrow +\infty$, we have
\begin{equation}
MSD(t\rightarrow +\infty)=(\frac{\pi^2-4}{\pi^3})L^2\approx 0.189 L^2.
\end{equation}
Because based on the Eq.(\ref{eq:I}) for the second term, the divergent part offsets the first term related to $ V(t)$, we obtain(with small $t$)
\begin{eqnarray}
MSD=PV\approx \frac{L^2}{2\pi \hat{Z}(L,t)}[(4\pi^{-3/2})\sqrt{A}]= \frac{4}{\pi}(2Dt)[1+\frac{2\sqrt{\pi Dt}}{L}].
\end{eqnarray}
\section*{VI. Simulation through random walk theory}
\begin{figure}
 \centering
      \subfloat{\includegraphics[width=0.8 \textwidth]{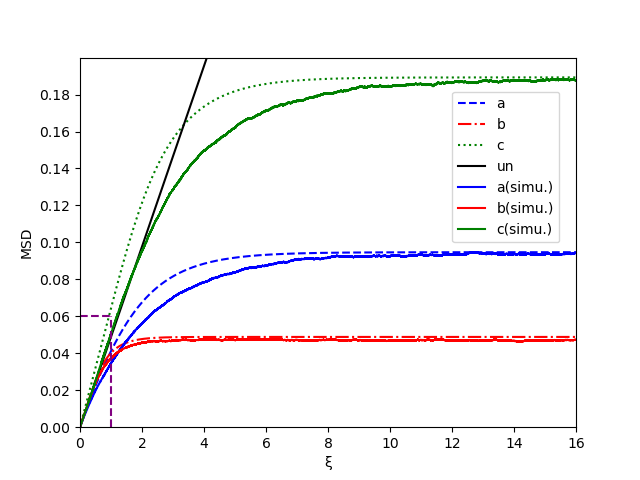}}

   \subfloat{\includegraphics[width=0.8 \textwidth]{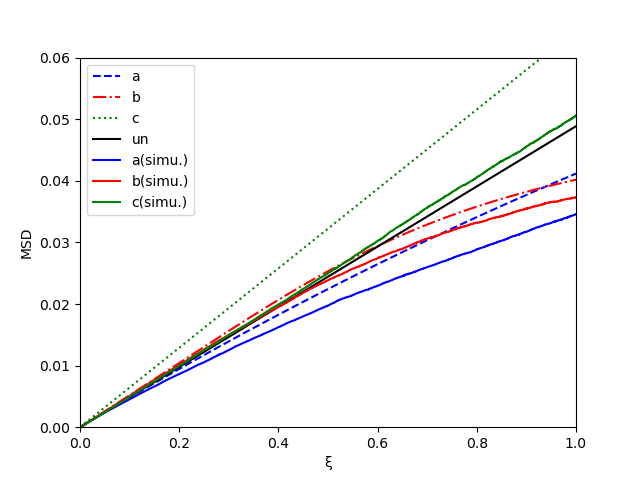}}
   \caption{Top panel:  the simulation result of MSD (in the unit of $\sqrt{\pi}N^2_m$) dependent of the reduced time $\xi$ ($\xi \equiv i/ i_{ch}$) with the diffusion coefficient $D=0.5$ for the initial condition (a),(b) and (c). Here $N_m=800$ and $i_m=960000$. There exists asymptotic value $0.095$, $0.047$ and $0.189$ in the case of large $\xi$ for the condition (a),(b) and (c), respectively. The blank solid line labelled by the symbol "un" corresponds the unconfined system where $MSD/(\sqrt{\pi}N_m^2)=0.0489\xi$ with the slope $0.0489$ decided via the limit behavior in small $\xi$ case for condition (a). Bottom panel: the simulation result of MSD dependent of small $\xi$ for the different initial condition (a), (b) and (c).}
 \label{fig:4}
\end{figure}
The above theoretical result is shown in Fig.(\ref{fig:3}). The Fokker-Planck equation could be derived by the random walk theory. The position reads $x_i$ for a particle which randomly takes $i$ steps ($i\in[0, i_m]$), with $x_i\in[0,N_m]$ for the confined Brownian motion. When $x_i\neq 0$ and $x_i\neq N_m$, $x_{i+1}=x_{i}\pm 1$, with the probability being $0.5$, respectively. When $x_i=0$ , $x_{i+1}=1$, When $x_i=N_m$, $x_{i+1}=N_m-1$. It means that $D=0.5$. We need to introduce a re-scaling relation where $ t\rightarrow i$, $L^2\rightarrow\sqrt{\pi}N^2_m$ and $t_{ch}\rightarrow i_{ch} \equiv\frac{\sqrt{\pi}N^2_m}{2\pi^2}.$ The simulation result is shown in Fig.(\ref{fig:4}). It need to be pointed that $x_i$ and $x_0$ is symmetrical under the condition (a) in the simulation. Therefore, $PV$ is equal to $PV(0)$ which satisfies $PV/(\sqrt{\pi}N_m^2)\approx \frac{1}{12\sqrt{\pi}}\approx 0.047$. For the confined system, there is some difference between Fokker-Planck equation and random walk theory, specially for the endpoint case (c).
\section*{VI. Results and discussion}
\label{sect:Results}
 Based on the series solution in Eq.(\ref{eq:MSD}), we obtain the MSD being $2Dt(1-\frac{2\sqrt{\xi} }{3\pi\sqrt{\pi}})$ for smaill $t$, with $t_{ch}=\frac{L^2}{4\pi^2D},\xi\equiv \frac{t}{t_{ch}}$, and the power $\alpha(t)$  being $\frac{1-0.18\sqrt{\xi}}{1-0.12\sqrt{\xi}}$. Further, as shown in Table.\ref{Tab:d}, we analysis the MSD and the power for the $d$-dimension system with $\gamma$-dimension confinement. In the case of $\gamma<d$, when $t$ is small or large enough, the diffusion is normal($MSD\propto t$). However, there exists the sub-diffusive behavior in the intermediate time. The universal description is consistent with the recent experiments and simulations in the micro-nano systems.

In the Ref.\cite{Ernst}, there is a foundational formula in previous researches for the confined system, which reads
\begin{eqnarray}
MSD_L(t)=\frac{L^2}{6}-\frac{16L^2}{\pi^4}\underset{m=0}{\overset{+\infty}{\sum}}\{[\frac{1}{(2m+1)^4}]\exp[-\frac{(2m+1)^2\pi^2D t}{L^2}] \}.
\end{eqnarray}
Here, $MSD_L(0)=0,MSD_L(t\rightarrow 0)\approx 2Dt$.  The formula has been widely used to discuss the diffusion of nano-materials,such as nanoporous structure\cite{Krager}. Under the condition $\bar{Z}=1$, the formula is very different of the  series solution in the Eq.(\ref{eq:MSD}). It is pointed out that it is similar with the PV when $\bar{Z}\approx 1$. We have
\begin{eqnarray}
PV&=&\frac{L^2}{2}-\frac{16L^2}{\pi^4}\underset{m=0}{\overset{+\infty}{\sum}}\{[\frac{1}{(2m+1)^4}]\exp[-\frac{(2m+1)^2\pi^2D t}{L^2}]\}-(\frac{L}{2})^2
\\ \nonumber &=&\frac{L^2}{12}+MSD_L(t).
\end{eqnarray}
Here, $PV(t=0)= \frac{L^2}{12}$. When the time $t$ is small, we have a formula being similar to he Eq.(\ref{eq:compare}), which reads
\begin{eqnarray}
PV(t)-PV(t=0)=2Dt(1-\frac{8\sqrt{Dt} }{3\sqrt{\pi}L}).
\end{eqnarray}
It also reflects the sub-diffusive behavior presented in the Ref.\cite{Ernst}. In previous studies, the MSD and the PV is equivalent to describe diffusion behavior. But in the paper we find that both is very different in the finite system. and we think that the Eq.(\ref{eq:MSD}) is a better choose to study all kinds of macro-nano systems.

 The initial conditions are essential to characterize the diffusion behavior described by the FPE, especially in the finite system. As examples, the paper discusses two representative initial PDFs reading $p_0(x)=\delta(x-x_0)$, with the midpoint $x_0=\frac{L}{2}$, and the endpoint $x_0=\epsilon$(or $0^+$). As shown in As the shown in Fig.(\ref{fig:3}),  In the case of midpoint, the MSD reads $2Dt(1-\frac{5\pi^3 Dt}{L^2})$ for the small $t$, which reflects a kind of sub-diffusion, with $D$ being the diffusion coefficient. In the case of endpoint, the MSD reads $\frac{4}{\pi}(2Dt)[1+\frac{2\sqrt{\pi Dt}}{L}]$ for the small $t$, which reflects a kind of super-diffusion. How to understand this type of super-diffusion behavior? We use the Dirichlet boundary  and also specify that the conservation of probability within the interval $L$. In a certain sense, the boundary is actually equivalent to a reflective boundary. There is a forced one-way diffusion initially which is faster than the normal diffusion.
\section*{Acknowledgments}
Y. Liao would thank Li-Cong Hu, Zi-Kai Fu, Xiao-Feng Zhou, Yu-Zhou Hao, Zhi-Bin Gao, Xiang-Ying Shen, Jian-Ying Du and Zi-Qian Xie for drawing assistance and writing embellishment. Y. Liao would be extremely appreciative of Prof. Bao-Wen Li for helpful discussion. This work was supported by National Natural Science Foundation of China (NSFC) under grants (42577094).

\end{document}